\documentclass[onecolumn, amsmath, amssymb, floatfix]{revtex4}

 \usepackage{amsmath,mathrsfs,graphicx,epstopdf,placeins,float}
 \usepackage{booktabs}
 \usepackage{multirow}
 \usepackage{}

\begin{document}

 \title{ Dynamic domain wall in charged dilaton black hole spacetimes}
 \author{Wu-Long Xu}
 \email{wlxu@emails.bjut.edu.cn}
 \author{Ai-Chen Li}
 \email{lac@emails.bjut.edu.cn}
 \author{Yong-Chang Huang}
 \email{ychuang@bjut.edu.cn}
 \affiliation{Institute of Theoretical Physics, Beijing University of Technology, Beijing 100124, China}
 \begin{abstract}
 In this paper we study the dynamics of $ n-1$ dimensional domain wall universe embedded in a $n$ dimensional charged dilaton black hole bulk with the non-asymptotically flat and non-asymptotically (A)dS characters.   We find that  domain wall can always  cross horizon,  for which  the  dilaton coupling constant $a$ and the ratio of pressure to density $w$ play the role of  controlling parameters. Then the consequent domain wall motion outside  the black hole generally falls into four situations: all time accelerating expansion, slow expansion with a constant speed, expansion followed by collapsing into horizon, and accelerating collapsing into horizon. However, there exists a small patch of  parametric sphere, in which the domain wall expansion first slows down and then accelerates. Our analysis also reveals that the big bang theory applies to the domain wall world scenario, while big bounce will not appear in our paper. Furthermore, we find that when we choose the expansion stage as radiation or matter stage, the resultant coupling strength   between  dilaton field and Maxwell field depends on the different black hole models.
 \end{abstract}
 \maketitle
\section{Introduction}

Dilaton gravity has a close relation with the low energy limit of string theory. The solution of charged dilaton black hole is gotten in the string theory \cite{Garfinkle:1990qj}. The corresponding thermodynamics of black hole and the relevant General relativity problems have been done. If the dilaton want to acquire mass in the low energy limit of string theory, for the way of obtaining mass, the literatures \cite{Gregory:1992kr},\cite{Horne:1992bi} have a discussion.

For the generalization of  charged dilaton black hole, it includes two ways. One is for cosmology constant term. One is for Maxwell field.
When the action exists the cosmology constant term,  Ref. \cite{Wiltshire:1994de} studies the  properties of  black holes and Ref. \cite{Poletti:1994ff}
studies the global properties of static spherically symmetric charged dilaton spacetimes. When one takes the dilaton potential as the  sum of Liouville potential, it can be solved a series of solutions of black hole \cite{Chan:1995fr},\cite{Sheykhi:2007wg}. All these solutions have unusual asymptotics character.
Furthermore, Ref. \cite{Gao:2005xv} gets the asymptotically flat and (A)dS solution of black hole with the sum of three Liouville potentials. For the Maxwell field, it can be generalized to nonlinear electrodynamics \cite{Hajkhalili:2018aqk},\cite{Sheykhi:2006dz}.

In the early universe, domain wall is a kind of topological defects considered as a product of vacuum phase transition. It can be evolved into stars such as  sheetlike or filamentary structures or voids \cite{Vilenkin:2000}. And in recent years, domain wall has always been the subject of research \cite{Vaquero:2018tib}-\cite{Matsui:2016xnp}.  When domain wall meets black hole, the interaction  among them have been studied \cite{Flachi:2006hw, Higaki:2000yv}. Beyond that, domain wall is also considered as a world volume in a high dimensional spacetime \cite{Rubakov:1983bb}-\cite{Lukas:1998yy}. Expansion or collapse of world volume can be reduced by the motion of domain wall in the bulk.  Thus  the  dynamics of domain wall in a bulk is very meaningful for understanding the evolution of universe \cite{Lee:2009sw}-\cite{Mazharimousavi:2011km}, on which H. S. Reall had done  significant work \cite{Chamblin:1999ya}-\cite{Chamblin:1999ea}.

In  general, the study of dynamics of domain wall in a  bulk classifies into two methods. One is to immediately obtain the dynamics of domain wall via the Israel matching conditions in an arbitrary background spacetime  \cite{Lee:2009sw},\cite{Maity:2008su}. The second way includes two steps due to the  bond between the action of domain wall and bulk metric (bulk spacetime is not  given). So at first one needs to find appropriate bulk metric solutions matching the given action of domain wall. Then the motion of domain wall can be analyzed with the junction conditions. As one case of domain wall-bulk bond, dilaton field  permeates  bulk and domain wall \cite{Chamblin:1999ya, Maity:2008rm, Mazharimousavi:2011km}. In this case, it needs to modify the dilaton coupling constants to get a static bulk metric meeting the field equations in the domain wall action.  And then  the evolution equation can be utilized to  get the  motion of domain wall.

The solutions of $n$ dimensional spacetime  in the Einstein-Maxwell-dilaton system with  the non-asymptotically flat and non-asymptotically (A)dS characters was proposed by K. C. K. Chan et al \cite{Chan:1995fr}. In those spacetimes, it  has a singularity located at the origin point. Based on these peculiarities, it can be extended to many meaningful researches  such as Thermodynamics \cite{Dehghani:2018svw,Sheykhi:2007wg}, Hawking Radiation \cite{SeyedehFatemehMirekhtiary:2015uia}, Holography \cite{Gouteraux:2011qh} et al. Now it needs to study the dynamics of domain wall under the non-asymptotically flat and non-asymptotically (A)dS background spacetimes assuming the matter of the domain wall is perfect fluid. It is desirable to find  the coupling strength  among dilaton field and Maxwell field,  and the ratio among density and pressure in the early universe, that can trigger the accelerating  expansion of the domain wall.

Our paper is arranged as follows: In  Section II we will simply review the non-asymptotically flat and non-asymptotically (A)dS background spacetimes. In Section III we will derive the dynamics equation of domain wall. In Section IV we will specifically analyze the motion of the domain wall in different spacetimes in  five dimensions. In  Section V we analyze the effect of dilaton field in the wall-bulk system. In  Section VI we give the conclusion and discussion.
\section{Charged dilaton black hole bulk}
In this section, we will simply review the background spacetimes for the convenience of  physical analysis. The relevant details can be found in the Ref. [22].

Let us  start from the action which could be derived  from the bosonic sector of the effective action in the IIA supergravity \cite{Chamblin:1999ea} in $n$-dimension Einstein-Maxwell-dilaton  bulk.
\begin{equation}\label{1}
S_{bulk}=\int d^nx\sqrt{-g}[R-\frac{4}{n-2}(\nabla \phi)^2-V(\phi)-e^{\frac{-4a\phi}{n-2}}F^2],
\end{equation}
where $\phi$ denotes the dilaton field, $V(\phi)$ is the dilaton potential and  $a$ is considered as  the coupling constant between dilaton field and Maxwell field. $F$ denotes the electromagnetic field tensor, $F_{\mu\nu}=\partial_{\mu}A_{\nu}-\partial_{\nu}A_{\mu}$.
Varying Eq. (\ref{1}) with respect to the metric $g^{AB}$, the Einstein field equation  is obtained as
\begin{equation}\label{2}
\begin{split}
R_{AB}=&\frac{4}{n-2}(\partial_A\phi\partial_B\phi
+\frac{1}{4}g_{AB}V)
+2e^{-\frac{4a\phi}{n-2}}(F_{AC}F^C_B\\
&-\frac{1}{2(n-2)}g_{AB}F^2).
\end{split}
\end{equation}
Maxwell and dilaton field dynamic equations are respectively given by
\begin{equation}
\partial_c(\sqrt{-g}e^{-\frac{4a\phi}{n-2}}F^{cd})=0,
\end{equation}
\begin{equation}\label{a}
\begin{split}
\frac{8}{n-2}\nabla^2\phi-\frac{\partial V(\phi)}{\partial\phi}+\frac{4a}{n-2}e^{\frac{-4a\phi}{n-2}}F^2=0.
\end{split}
\end{equation}
The metric of static spacetime is written as
\begin{equation}\label{4}
ds^2=-U(r)dt^2+\frac{1}{U(r)}dr^2+R^2(r)d\Omega_{n-2}.
\end{equation}
Correspondingly, the  Maxwell field is an isolated electric charge, which is
\begin{equation}\label{5}
F_{tr}=e^{\frac{4a\phi}{n-2}}\frac{Q}{R^{n-2}},
\end{equation}
where $Q$ is the charge of black hole. Submitting Eqs. (\ref{4}) and (\ref{5}) into Eqs. (\ref{2}-\ref{a}), one obtains
\begin{eqnarray}
&&\frac{1}{R}\frac{d^2R}{dr^2}=-\frac{4}{(n-2)^2}(\frac{d\phi}{dr})^2,\\
&&\frac{1}{R^{n-2}}\frac{d}{dr}(R^{n-2}U\frac{d\phi}{dr})=\frac{n-2}{8}\frac{dV}{d\phi}+a e^{\frac{4a\phi}{n-2}}\nonumber\\
&&~~~~~~~~~~~~~~~~~~~~~~~~~~~~~~~\times\frac{Q^2}{R^{2(n-2)}},\\
&&\frac{1}{R^{n-2}}\frac{d}{dr}[U\frac{d}{dr}(R^{n-2})]=(n-2)(n-3)\frac{1}{R^2}-V\nonumber\\
&&~~~~~~~~~~~~~~~~~~~~~~~~~~~~~~~-2e^{\frac{4a\phi}{n-2}}\frac{Q^2}{R^{2(n-2)}}.
\end{eqnarray}

Similar to Ref. \cite{Chan:1995fr}, we consider the metric ansatz
\begin{equation}
R(r)=\gamma r^N,
\end{equation}
where $\gamma$ and $N$ are constants. With the aid of Eq. (7), the solution of $\phi$ is reached,
\begin{equation}
\phi(r)=\phi_0+\phi_1 \log r.
\end{equation}

For explaining physics, it is sufficient to consider two types of potential  and  $n=5$ in the following context 
\begin{align}
&V(\phi)=0\\
&V(\phi)=2\Lambda e^{2b\phi}
\end{align}
where $\Lambda$ is the cosmology constant and $b$ is the coupling constant between the dilaton field and cosmology constant (for the results of other potentials, see Ref. [22]). It gets four kinds of spacetime solutions (One is non-asymptotically flat and the others is non-asymptotically (A)dS).

Type I:  $V(\phi)=0$

The solution is as follows
\begin{align}
&U(r)=\frac{(4+a^2)^2}{\gamma^2(2+a^2)}r^{\frac{8}{4+a^2}}
-\frac{4(4+a^2)}{3a^2\gamma^{3}}Mr^{\frac{2(2-a^2)}{4+a^2}},\\
&N=\frac{a^2}{4+a^2},\\
&r_h=(\frac{4M(2+a^2)^2}{3a^2(4+a^2)\gamma})^{\frac{4+a^2}{2(2+a^2)}},\\
&\phi(r)=-\frac{3}{4a}\log{\frac{2Q(2+a^2)}{12\gamma^4}}+\frac{3a}{(4+a^2)}\log r,
\end{align}
where $M$ is the mass of the black hole and $r_h$ is the horizon radius of black hole. When the potential vanishes, the metric solution $U(r)$ shows that  the bulk will become a non-asymptotically flat spacetime.

Type II:  $V(\phi)=2\Lambda e^{2b\phi}$

(i) The first solution is as follows
\begin{eqnarray}
&&U_1(r)=r^{\frac{2a^2}{1+a^2}}[\frac{2(1+a^2)^2}{(1-a^2)\gamma^2(2+a^2)}-\frac{4(1+a^2)M}{3\gamma^3}\nonumber\\
&&~~~~~~~\times r^{\frac{-(2+a^2)}{1+a^2}}\!+\frac{2Q^2(1+a^2)^2}{3(2+a^2)\gamma^{6}}e^{\frac{4a\phi_0}{3}}r^{-\frac{2(2+a^2)}{1+a^2}}],\\
&&N=\frac{1}{1+a^2},\\
&&\phi_1=-\frac{3a}{2(1+a^2)},\\
&&\Lambda=-\frac{3a^2}{\gamma^2(1-a^2)}e^{-\frac{4\phi_0}{3a}},\\
&&b=\frac{2}{3a}.
\end{eqnarray}
When $a^2<1$, this spacetime has two horizons;  when $a^2>1$, it just has one horizon.

(ii) The second solution is as follows
\begin{eqnarray}
&&U_2(r)=r^{\frac{8}{4+a^2}}\{\frac{(4+a^2)^2}{(2+a^2)(4-a^2)}\times[-1+\frac{2Q^2}{2\gamma^{4}}e^{\frac{4a\phi_0}{3}}]\nonumber\\
&&~~~~~~~-\frac{4(4+a^2)M}{3a^2\gamma^3}r^{\frac{-2(2+a^2)}{4+a^2}}\},\\
&&N=\frac{a^2}{4+a^2},\\
&&\phi_1=\frac{3a}{4+a^2},\\
&&\Lambda=\frac{12}{(4-a^2)\gamma^2}e^{\frac{2a\phi_0}{3}}-\frac{2(2+a^2)Q^2}{(4-a^2)\gamma^6}e^{2a\phi_0},\\
&&b=-\frac{a}{3}.
\end{eqnarray}
In this situation, when $a^2>4$, the black hole shows up as a naked singularity. There exists one horizon when the right side of  $r_h$ is positive, namely $a^2<4$.

(iii) The third solution is as follows
\begin{eqnarray}
&&U_3(r)=\frac{(4+a^2)^2}{\gamma^2(2+a^2)}r^{2\frac{8}{4+a^2}}-\frac{4(4+a^2)M}{3\gamma^2a^2}r^{\frac{2(2-a^2)}{4+a^2}}\nonumber\\
&&~~~~~~~~+\frac{2\Lambda Q^{\frac{4}{a^2}}2^{\frac{2}{a^2}}[4+a^2]^2(2+a^2)^{\frac{2}{a^2}}}{a^2\gamma^{\frac{8}{a^2}}[4-4a^2]3^{\frac{2+a^2}{a^2}}2^{\frac{4}{a^2}}}r^{\frac{2a^2}{4+a^2}},\\
&&N=\frac{a^2}{4+a^2},\\
&&\phi_1=\frac{3a}{(4+a^2)},\\
&&\phi_0=-\frac{3}{4a}\log\frac{Q^2(2+a^2)}{6\gamma^4},\\
&&b=-\frac{2}{a}.
\end{eqnarray}
It is pretty difficult to define black hole horizon. Because when we take $U(r)=0$,  $r$ takes multiple values  depending on the range of  constant $a$.
\section{dynamic equation of domain wall }
 In this section we will derive the dynamic equation of domain wall in the charged dilaton black hole background. Before that, it is necessary  to introduce some useful  notations about the domain wall-bulk system.  In general, $X^A=(t,r,X,Y,Z)$ denotes bulk coordinate and $x^\mu(\tau,x,y,z)$ denotes internal coordinate of domain wall, for a four dimensional  domain wall embedded parallelly into a five dimensional  bulk spacetime along the additional spacial dimension $r$. Then we have $X^A(x^{\mu})=X^A(t(\tau),r(\tau),x,y,z)$ -- $r(\tau)$ being the position of domain wall in the bulk. The domain wall metric will exist in the range $r<r(\tau)$  and the unit normal points to the range.  Meanwhile, we consider that the domain wall has the reflection symmetry. Thus we always take the value of $\{\}_{-}$ to calculate result. Also we  shuould know a tensor has a transforamble relation among the two different coordinates $T_{\mu\nu}=\frac{\partial X^M}{\partial x^{\mu}}\frac{\partial X^N}{\partial x^{\nu}}T_{MN}$. 

 After this, our calculations start from defining a four vector followed \cite{Brax:2002nx}
\begin{equation}
e^{A}_{\mu}=\frac{\partial X^A}{\partial x^{\mu}},
\end{equation}
then, the induced metric in the domain wall is
\begin{equation}
h_{\mu\nu}=g_{AB}e^A_{\mu}e^B_{\nu}.
\end{equation}
When we take the general static bulk metric ansatz
\begin{equation}
dS^2=-A(r)dt^2+B(r)dr^2+R(r)^2d\Omega^2,
\end{equation}
the velocity of the domain wall is
\begin{equation}
V^A=(\sqrt{\frac{B\dot{r}^2+1}{A}},\dot{r},0,0,0),
\end{equation}
where a dot donates the derivative with respect to $\tau$ and we use $g_{AB}V^AV^B=-1$. The normal vector is
\begin{equation}
n^A=(-\sqrt{\frac{B}{A}}\dot{r},-\sqrt{\frac{1+B\dot{r}^2}{B}},0,0,0),
\end{equation}
where we use the conditions:
\begin{align}
&g_{AB}n^An^B=1,\\
&n^AV_A=0.
\end{align}
And we consider the wall as the FRW university. The metric is
\begin{equation}
dS^2=-d\tau^2+R^2(r(\tau))d\Omega^2.
\end{equation}
The extrinsic curvature tensor is
\begin{equation}
K_{\mu \nu}=e^A_{\mu}e^B_{\nu}K_{AB}=e^A_{\mu}e^B_{\nu}\nabla_A n_B.
\end{equation}
Explicitly, the extrinsic curvature components are \cite{Visser:1995cc}
\begin{align}
&K_{ij}=-\frac{h_{ij}}{R}R'\sqrt{\frac{1+B\dot{r}^2}{B}},\\
&K_{\tau\tau}=\frac{1}{\sqrt{AB}}\frac{d}{dr}(\sqrt{A(1+B\dot{r}^2)}).
\end{align}

The geometry of domain wall and the background spacetime is known clearly now.  How the domain wall moves in the bulk can be identified by the Israel matching conditions. For that, firstly we introduce the $n-1$ dimensional domain wall action
\begin{equation}
S_{DW}=\int d^{n-1}x\sqrt{-h}\{K\}+\int d^{n-1}x\sqrt{-h}L_m(\phi, \tilde{h}_{\mu\nu}),
\end{equation}
where we adopt an assumption about the induced metric $h_{\mu\nu}$ \cite{Brax:2002nx}
\begin{equation}
\tilde{h}_{\mu\nu}=e^{2\xi(\phi)}h_{\mu\nu},
\end{equation}
and  consider the matter in the domain wall as perfect fluid. So the energy-momentum tensor can be
\begin{equation}
 T^{\mu\nu}=(\rho+p)V^{\mu}V^{\nu}+ph^{\mu\nu},
\end{equation}
where $\rho$, $p$ are density  and  pressure of the perfect fluid correspondingly.

Varying the total action $S=S_{bulk}+S_{DW}$ with respect to  $g^{AB}$,  the $n-1$ dimensional  field equation is
\begin{equation}\label{Israelequation1}
\{K_{AB}-Kh_{AB}\}=\tau_{AB},
\end{equation}
where $\tau_{AB}=-2\frac{\delta L_m}{\delta h^{AB}}+h_{AB}L_m$. Thus we gain the Israel matching conditions \cite{Israel:1966rt}.

And varying with  dilaton field  $\phi$, the scalar field equation is
\begin{equation}\label{scalarequation1}
\frac{4}{n-3}\{n\cdot\partial\phi\}=\frac{dL_M}{d\phi}.
\end{equation}
So, there are three boundary equations from the Eq. (\ref{Israelequation1}) and Eq. (\ref{scalarequation1})
\begin{align}
&K_{ij}=-\frac{h_{ij}}{R}R'\sqrt{\frac{1+B\dot{r}}{B}}=-\frac{1}{2}(\tau_{ij}-\frac{1}{3}\tau h_{ij}),\\
&K_{\tau\tau}=\frac{1}{\sqrt{AB}}\frac{d}{dr}(\sqrt{A(1+B\dot{r}^2)})=-\frac{1}{2}(\tau_{\tau\tau}-\frac{1}{3}\tau h_{\tau\tau}),\\
&\frac{8}{n-3}n\cdot\partial\phi=\xi'\tau.
\end{align}
When $n=5$, we can expand  Eq. (\ref{2}), it will have
\begin{equation}
\frac{d}{dr}(\frac{R}{R'})=1+\frac{4R^2\phi'^2}{9R'^2}.
\end{equation}
Using metric (\ref{4}) and energy-momentum tensor $T^{\mu\nu}$, we will deform boundary equations as a new face
\begin{align}
&\sqrt{U(r)+\dot{r}^2}=\frac{R}{R'}\frac{1}{6}\rho,\\
&\phi'=\frac{9}{4}\frac{R'}{R}\xi'(3w-1),\label{9}\\
&\dot{\rho}+3H(P+\rho)=-\frac{4}{9}\frac{\rho\phi'^2}{H},
\end{align}
where the $w=\frac{P}{\rho}$, $H=\frac{\dot{R}}{R}$.

We have $R(r)=\gamma r^{N} ,\phi(r)=\phi_{0}+\phi_{1}\log r$. Solving (\ref{9}), we get the expression of $\xi(\phi)$
\begin{equation}
\xi(\phi)=\frac{4}{9N(3w-1)}\phi+\phi_2.
\end{equation}
The Friedmann equations from the remaining boundary equations have the form
\begin{align}
&H^2=\frac{1}{36}\rho^2-\frac{N^2}{r^2}U(r),\label{10}\\
&\dot{\rho}+3H(P+\rho)=-\frac{4}{9}\rho\phi'^2H(\frac{R}{R'})^2\label{11}.
\end{align}
Solving Eqs. (\ref{10}) and (\ref{11}),  we can get the analytical expressions
\begin{align}
&\rho=\rho_1\frac{1}{\gamma}R^{-(3+3w+\frac{4\phi_1^2}{9N^2})},\\
&\dot{R}^2-Y(R)=0,
\label{12}
\end{align}
where $Y(R)=\frac{1}{36}(\rho_1R^{-(2+3w+\frac{4\phi^2_1}{9N^2})})^2-R^2\frac{N^2}{r^2}U(r)$.

From  Eq. (\ref{12}), we  know that it decides the evolution of the universe scale factor. In physics, it shows a conversion process of energy  from the potential $Y(R)$ to kinetic energy for a particle.
\section{domain wall evolution}
Before we analyze the motion of domain wall based on the evolution  equation for scale factor, we need to do some skillful analysis. From the Eq. (\ref{12}), it can be known easily that $Y(R)$ should be big than 0 if this expression holds. While  in the range  $0<R<1$ and $1<R<\infty$, $Y(R)$ has reverse monotonicity. Thus it is sufficient to  analyze the range $1<R<\infty$ by scanning the all parameters $a$, $w$ for finding the situation of $Y(R)>0$.  Then according to the monotonicity of $Y(R)$, the domain wall position $R_0$ in the point of $\dot{R}=0$ and the corresponding horizon radius $R_h$,  the motion of domain wall will be clear outsider of the horizon. Also the physics corresponding to the domain wall being inside the horizon or  going across the horizon  is not known for us. Thus it only can  understanded  the laws of motion of domain wall  out of horizon. Then in the Eq. (\ref{12}),  obviously we can know from the expression of Y(R) when $\rho_1\rightarrow \infty$, $Y(R)>0$. Domain wall can always expand.  Finally, we take general situation $\gamma=1,\rho_1=1,\phi_0=0$. This would does not change property of $Y(R)$.

Type I: solution

The metric $U(r)$ is
\begin{equation}
U(r)=\frac{(4+a^2)^2r^{\frac{8}{4+a^2}}}{(2+a^2)^2}
-\frac{4(4+a^2)M}{3a^2}r^{\frac{2(2-a^2)}{4+a^2}}.
\end{equation}
According to the monotony of  $U(r)$, there is only one  horizon. At the same time, because its  power is higher than 0, when $r\rightarrow \infty$  this bulk spacetime is non-asymptotically flat.

Then it has
\begin{equation}
\begin{split}
Y(R)=&\frac{4a^2M}{3(4+a^2)}R^{-\frac{4}{a^2}-2}+\frac{\rho_1^2}{36}R^{-4-6w-\frac{8}{a^2}}\\
&-\frac{a^4}{(2+a^2)^2}.
\end{split}
\end{equation}

For simplicity, we can take the form of $Y(R)$
\begin{equation}
Y(R)=R^{C_1}+R^{C_2}-C_3.
\end{equation}

We can directly see that $C_1<-2$. If $w>-\frac{2}{3}$,  then $C_2<0$.  and then $Y(R)$ decreases monotonously. Because of the constant term is less than 0. When $R\rightarrow\infty$, $Y(R)<0$.

$C_2>0$, then $w<\frac{-2}{3}-\frac{4}{3a^2}$. When $a^2<1$, $w$ doesn't have a solution. When $a^2>1, w<-3$, $Y(R)$ will be dominated by $R^{C_2}$. So Y(R) increase monotonously.

Case (I).  For $C_2<0$, $w>-\frac{2}{3}$. As shown in FIG. 1, $Y(R)$ will quickly decrease to 0. It has nothing to do with value of $a$. In this situation,  the domain wall position $R_0$ in the point of $\dot{R}=0$ and the corresponding horizon radius $R_h$. $R_0>R_h$ shows that the domain wall  expands followed by collapsing into horizon.
\begin{figure}[H]
\center
  \includegraphics[width=4.2cm]{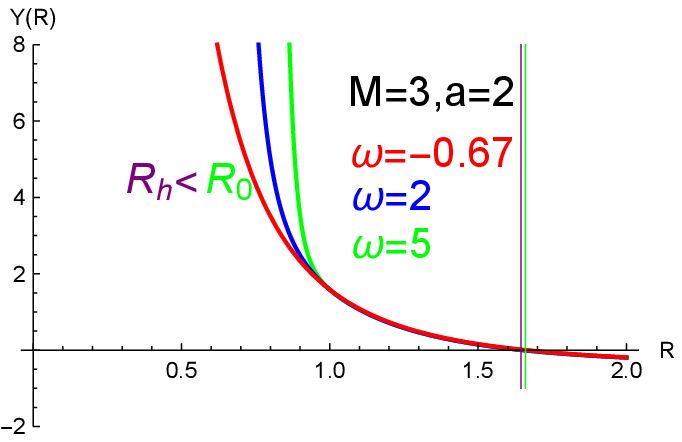}
  \includegraphics[width=4.2cm]{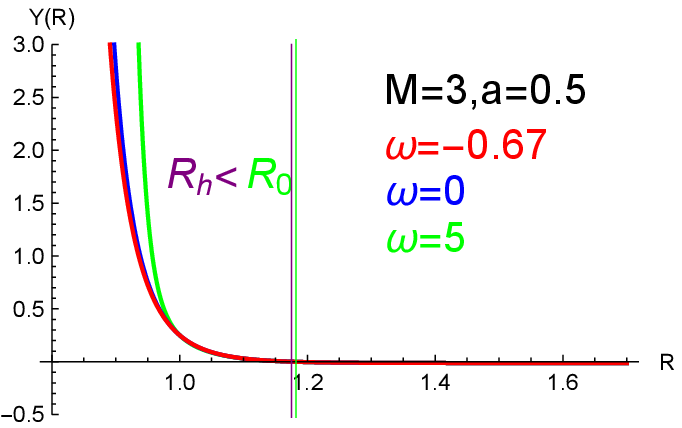}\\
  \caption{$Y(R)$ versus $R$ for the Type I's solution.}
\end{figure}

Case (II). For $a^2>1, w<-3$, $Y(R)$ fastly increases. From FIG. 2, when $\dot{R}>0$, the domain wall moving is accelerating expansion all time. When $\dot{R}<0$, the domain wall  collapses into horizon quickly.
\begin{figure}[H]
\center
  \includegraphics[width=6cm]{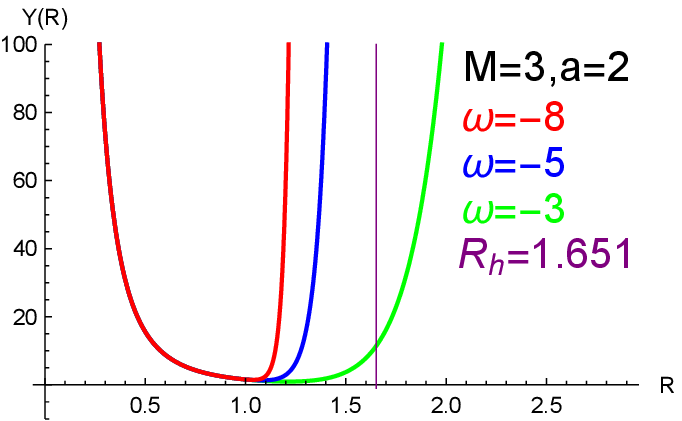}\\
  \caption{$Y(R)$ versus $R$ for the Type I's solution. }
\end{figure}

Type II: solutions

(I)The metric $U(r)$ is
\begin{eqnarray}
&&U(r)=\frac{2(1+a^2)^2e^{\frac{4a\phi_0}{3}}Q^2}{3(2+a^2)}r^{\frac{-4}{1+a^2}}
-\frac{4}{3}(1+a^2)Mr^{\frac{-2}{1+a^2}}\nonumber\\
&&~~~~~~~+r^{\frac{2a^2}{1+a^2}}\frac{2(1+a^2)^2}{(1-a^2)(2+a^2)},\\
&&\Lambda=\frac{-3a^2}{1-a^2}e^{\frac{-4\phi_0}{3a}}.
\end{eqnarray}
From this bulk solution, we know $a\neq1, a>0$. And this spacetime is non-flat and non-(A)dS. When $a^2<1, \Lambda<0$, and $a^2>1, \Lambda>0$, we have
\begin{equation}
\begin{split}
Y(R)=&-\frac{2Q^2R^{-2(2+a^2)}}{3(2+a^2)}+\frac{4MR^{-2-a^2}}{3(1+a^2)}
+\frac{\rho_1^2}{36}R^{-4-2a^2-6w^2}\\
&-\frac{2}{(1-a^2)(2+a^2)}.
\end{split}
\end{equation}
Similarly, for simpleness, we take the form of $Y(R)$
\begin{equation}
Y(R)=-R^{C_1}+R^{C_2}+R^{C_3}+C_4.
\end{equation}
It can be seen  $C_1<C_2<0$, for which $R^{C_1}<R^{C_2}<1$. So the first two terms of $Y(R)$ are dominated by $R^{C_2}$.

For $\Lambda<0, a^2<1$. If $C_3>0$, then $w<-1$. $Y(R)$ will be dominated by $R^{C_3}$ and  increases monotonously. If $C_3<0, w>-\frac{2}{3}$, $Y(R)$  decreases monotonously. Meanwhile, due to that $C_4$ is very small, $Y(R)$ must be less than 0 quickly.

For $\Lambda>0, a^2>1$. If $C_3>0$, then $w<\frac{-2-a^2}{3}$.  We cannot define the value of $Y(R)$. If $C_3<0$, then $w>-1$. $Y(R)$  decreases monotonously. Although $Y(R)$ decrease all time, because of $C_4>0$, $Y(R)$ eventually closes into a constant.

Case (I). For $\Lambda<0, a^2<1,w<-1$, in FIG. 3 $Y(R)$ fastly increases. When $\dot{R}>0$, the  domain wall is accelerating expansion. And it can't stop. When $\dot{R}<0$, the domain wall collapses immediately into horizon.
\begin{figure}[H]
\center
  \includegraphics[width=6.0cm]{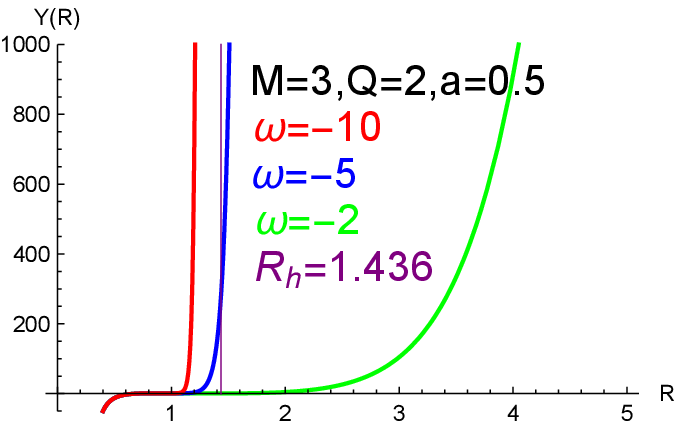}\\
  \caption{$Y(R)$ versus $R$ for the first Type II's solutions. }
\end{figure}
Case (II). For $\Lambda>0, a^2>1$. If $w>-1$, this situation is interesting. When $\dot{R}>0$, domain wall will expand in all the time. And its velocity closes to a constant eventually. The other is domain wall collapses into horizon.
\begin{figure}[H]
\center
  \includegraphics[width=6cm]{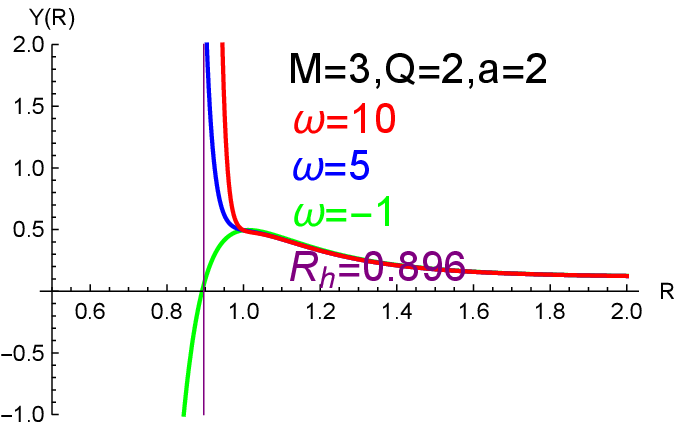}\\
  \caption{$Y(R)$ versus $R$ for the first Type II's solutions. }
\end{figure}

Case (III). For $a^2<1$, $w>-\frac{2}{3}$. $Y(R)$ will have a form of inverted parabolic. But we calculate the domain wall position $R_0$ in the point of $\dot{R}=0$ and the corresponding horizon radius $R_h$. $R_0>R_h$ shows that the domain wall crosses the horizon followed by slow down expanding and  then collapses into horizon. The bounce will not appear. 
\begin{figure}[H]
	\center
	\includegraphics[width=6cm]{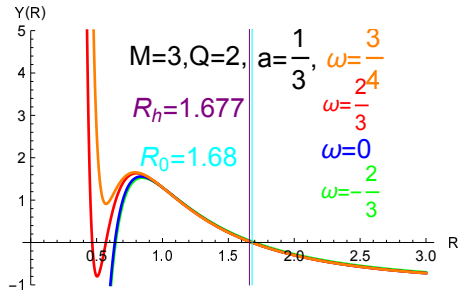}
	\caption{$Y(R)$ versus $R$ for the first Type II's solutions. }
\end{figure}
(II) The metric $U(r)$ is
\begin{eqnarray}
&&U(r)=r^{\frac{8}{4+a^2}}(\frac{(4+a^2)^2(-1+e^{\frac{4a\phi_0}{3}}Q^2)}{(4-a^2)(2+a^2)}\nonumber\\
&&~~~~~~~-\frac{4(4+a^2)Mr^{\frac{-2(2+a^2)}{4+a^2}}}{3a^2}),\\
&&\Lambda=\frac{-2(2+a^2)e^{2a\phi_0}Q^2}{(4-a^2)}+\frac{12e^{-\frac{2a\phi_0}{3}}}{4-a^2}.
\end{eqnarray}

In this situation, $a\neq2$. And just when $a^2<4$, the black hole has a black hole. Or it is just a naked singularity.  This bulk solution is non-flat and non-(A)dS. Because of the arbitrariness of $Q$, $\Lambda$ is not ensured.

\begin{equation}
\begin{split}
Y(R)=&\frac{4a^2M}{3(4+a^2)}R^{-\frac{2(2+a^2)}{a^2}}+\frac{\rho_1^2}{36}
R^{-2-\frac{2(4+a^2)}{a^2}-6w}\\
&-\frac{(-1+Q^2)a^4}{(4-a^2)(2+a^2)}.
\end{split}
\end{equation}

For analyzing,
\begin{equation}
Y(R)=R^{C_1}+R^{C_2}+C_3.
\end{equation}

Obviously, $C_1<0$. When $a^2>4, C_3>0$. If $C_2>0, then w<-1$. Y(R) is dominated by $R^{C_2}$ and  increases monotonously. If $C_2<0$, then $w>-\frac{2}{3}$. Y(R) decreases monotonously.  $Y(R)>0$ and  closes to a constant finally.

When $a^2<4, C_3<0$. If $C_2>0$, we cannot have a solution. If $C_2<0$, then $w>-1$. $Y(R)$  decreases monotonously. And $Y(R)<0$ quickly.

Case (I). For $a^2>4$, the bulk is a naked singularity.  If $w<-1$, the domain wall moving has two situations. One is slow expansion then accelerating expansion. One collapses into the singularity.

If $w>-\frac{2}{3}$, the domain wall is slowing  expansion  until velocity is a constant or collapses  into an horizon.

\begin{figure}[H]
\center
  \includegraphics[width=4.2cm]{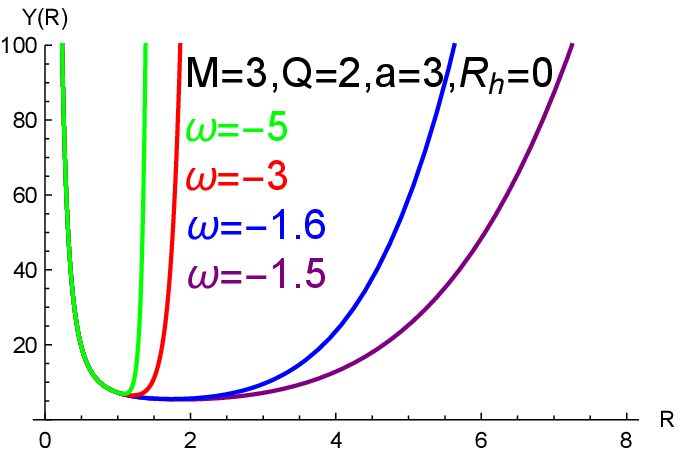}
  \includegraphics[width=4.2cm]{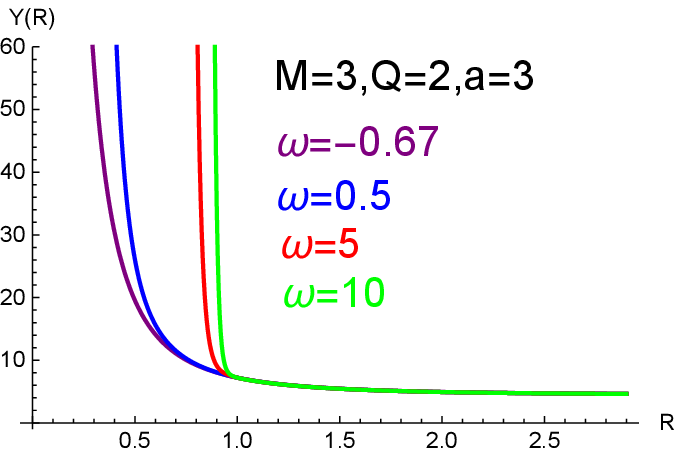}\\
  \caption{$Y(R)$ versus $R$ for the second Type II's solutions. }
\end{figure}

\begin{figure}[H]
	\center
	\includegraphics[width=4.2cm]{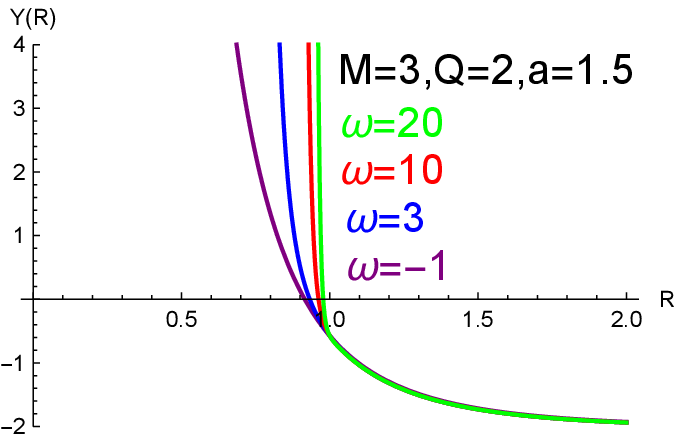}
	\caption{$Y(R)$ versus $R$ for the second Type II's solutions. }
	\label{123}
\end{figure}

Case (ii). In this spacetime, when $Y(R)$ decreases monotonously,  it can be observed by the FIG. (\ref{123}) that the domain wall slowly expands followed by collapsing into horizon or collapsing into horizon directly.

(III) The metric $U(r)$ is

\begin{equation}
\begin{split}
U(r)=&\frac{(4+a^2)^2}{(2+a^2)^2}r^{\frac{8}{4+a^2}}-\frac{2(4+a^2)}{a^2}Mr^{\frac{2(2-a^2)}{4+a^2}}\\
&+\frac{2^{1-\frac{2}{a^2}}3^{-\frac{2+a^2}{a^2}}(2+a^2)^{\frac{2}{a^2}}(4+a^2)^2Q^{\frac{4}{a^2}}}{a^2(4-4a^2)}
r^{\frac{2a^2}{4+a^2}}\Lambda.
\end{split}
\end{equation}

It shows that $a\neq1$ and this spacetime is non-flat and non-(A)dS due to the effect of $a$. At the same it is not sure the situation of horizon.

\begin{equation}\label{33}
\begin{split}
Y(R)=&-\frac{a^4}{(2+a^2)^2}+\frac{2(4+a^2)M}{a^2}R^{-2+\frac{4}{a^2}}\\
&-\frac{2^{-1-\frac{2}{a^2}}3^{-\frac{2+a^2}{a^2}}(2+a^2)^{\frac{2}{a^2}}(4+a^2)Q^{\frac{4}{a^2}}\Lambda}{a^2(1-a^2)}
R^2\\
&+\frac{\rho_1^2}{36}R^{-4-\frac{8}{a^2}-6w}.
\end{split}
\end{equation}
For analyzing and simplification, we don't consider  the situation of  limit of $\Lambda$.
\begin{equation}
Y(R)=R^{C_1}-mR^{2}+R^{C_2}-C_3.
\end{equation}

(a) $\Lambda>0$

 When $a^2>1, -2<C_1<2, m<0, R^{C_1}<R^{2}$. So the first two terms of $Y(R)$ will be dominated by the second term. If $C_2<0$, then $w>-\frac{2}{3}$. $Y(R)$ is dominated by $R^2$ and increases. If $C_2>0$, then $w<-2$. $Y(R)$  increases monotonously.

 When $0<a^2<1,C_1>2,m>0$. So the first two terms of $Y(R)$ is dominated by the first term. If $C_2<0$, then $w>-2$. $Y(R)$ is dominated by $R^{C_1}$ and increases monotonously.  For $C_2>0$, if $C_2>C_1$, it doesn't have a solution.  $C_2<C_1$, and then $w>-\frac{7}{3}$. $Y(R)$ is dominated by $R^{C_1}$ and increases monotonously.

(b) $\Lambda<0$

If $a^2>1$, then $-2<C_1<2, m>0$, and $R^{C_1}<R^{2}$.  So the first two terms of $Y(R)$ is dominated by the second term. So it just when $C_2>2, w<-\frac{7}{3}$, Y(R) is dominated by $R^{C_2}$. $Y(R)$ will increase monotonously. In fact when $C_2<2, w>-1$, we consider that if it exists a inverted parabolic picture for $Y(R)$. If it appears, it shows that this universe can process a periodic oscillation.

If $0<a^2<1, C_1>2, m<0$,  the first two terms of $Y(R)$ is dominated by the $R^{C_1}$. If $C_2<0$, then $ w>-2$. $Y(R)$ is dominated by $R^{C_1}$ and increases monotonously. For $C_2>0$, if $C_2>C_1$, $w$ not has a solution.  If $C_2<C_1$, then $w>-\frac{7}{3}$. $Y(R)$ is dominated by $R^{C_1}$ and increases monotonously.

Case (I). It shows that for $\Lambda>0$, $a^2>1, w>-\frac{2}{3}$, when $w$ increases, $Y(R)$ continually increases. The domain wall has two situations. One is slow expansion then  accelerating expansion, and one is collapsing into horizon.  When $a^2<1, w>-\frac{7}{3}$, the domain wall also has two situations. One is slow expansion then accelerating expansion, the other is collapsing into horizon. And the last expansion accelerates quickly very much.
\begin{figure}[H]
\center
  \includegraphics[width=4.2cm]{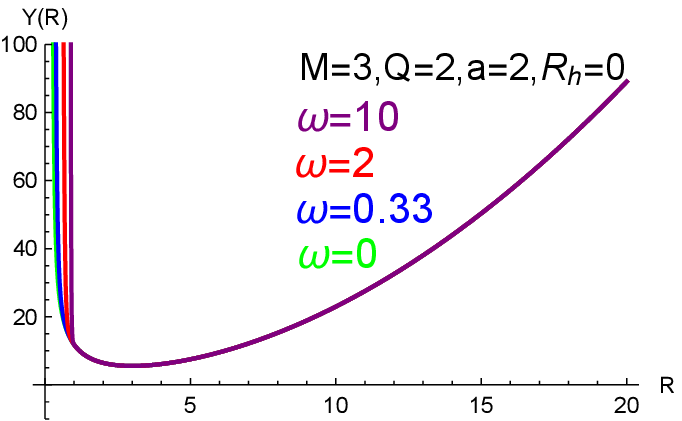}
  \includegraphics[width=4.2cm]{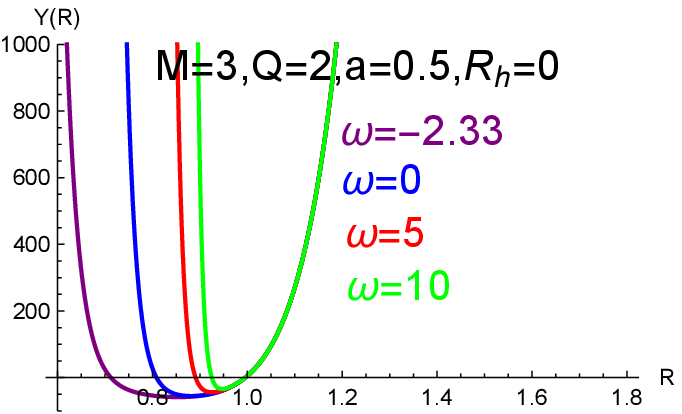}\\
  \caption{$Y(R)$ versus for the third Type II's solutions, $\Lambda=1$. Left picture is for $a^2>1$. Right picture is for $a^2<1$.}
\end{figure}
Case (II). For $\Lambda<0$, if $a^2>1$, then $w<-\frac{7}{3}$. When $w$ continually decreases, $Y(R)$ increases more and more faster. The motion of domain wall has two situations. One is accelerating expension. One is collapsing into horizon. If $a^2<1, w>-\frac{7}{3}$, domain wall has two situations. One is always accelerating expansion, the other is collapsing into horizon.
\begin{figure}[H]
\center
  \includegraphics[width=4.2cm]{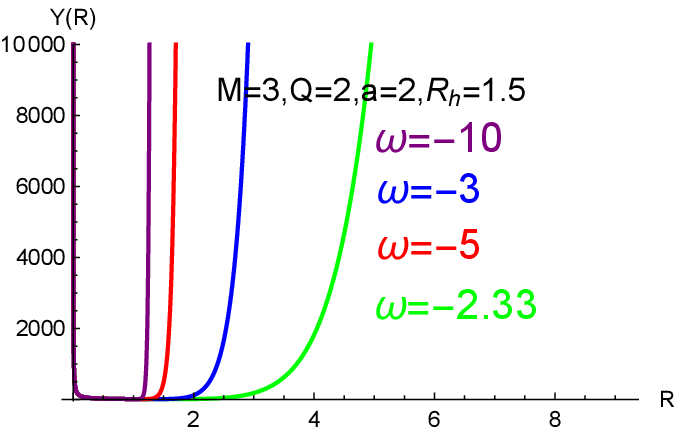}
  \includegraphics[width=4.2cm]{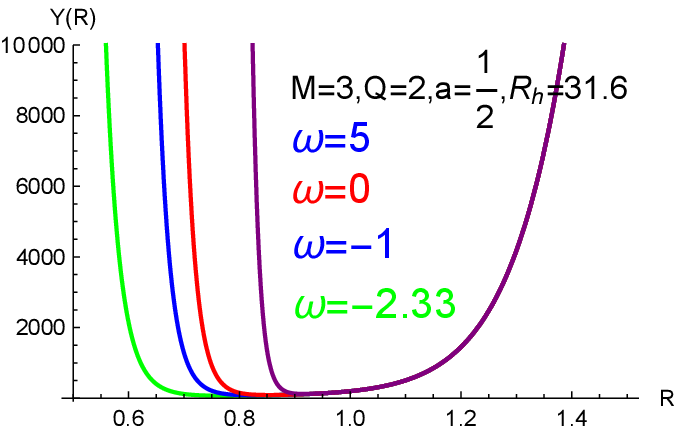}\\
  \caption{$Y(R)$ versus $R$ for the third Type II's solutions, $\Lambda=-1$.}
\end{figure}
Case (III).  For $\Lambda<0$, if $a^2>1, C_2<2, w>-1$, it shows that the inverted parabolic of $Y(R)$ doesn't appear from the FIG. (\ref{picture}), although $Y(R)$ exists $-R^2$. Thus in this metric solution the motion of  domain wall will not produce a periodic oscillation.
\begin{figure}[H]
\center
  \includegraphics[width=6cm]{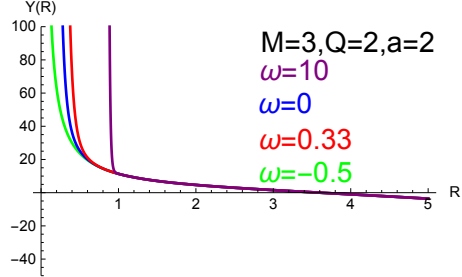}
  \caption{$Y(R)$ versus $R$ for the special Type II's solutions, $\Lambda=-1$.}\label{picture}
\end{figure}

Case (IV). By analyzing, there is a situation for Eq. (\ref{33}), in which  $Y(R)$ don't increase in all time if we adjust other parameters such as $M$, $Q$ or $\Lambda$. So that the domain wall is acceleraing expansion then  slows down the velocity of  expansion  into a minimum value and then starts accelerating expansion.
\begin{figure}[H]
\center
  \includegraphics[width=6cm]{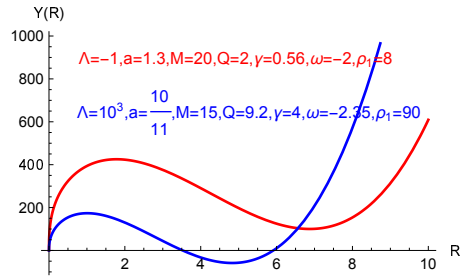}
  \caption{$Y(R)$ versus $R$ for the special Type II's solutions.}
  \label{11picture}
\end{figure}

Now we summarize that by altering the coupling constant $a$ and state parameter $w$, domain wall always can cross horizon. Outside of horizon,  there are four kinds of different evolution experiences for the domain wall. The first is  domain wall undergoes accelerating expansion all  time. The second is  expansion with a constant velocity. The third is slow expansion followed by collapsing into horizon. The fourth is accelerating collapsing into horizon. However, there exists a small patch of  parametric sphere, in which the domain wall expansion is accelerating  then  slows down and then accelerates.

Beyond that, it infers that all $Y(R)$ other than eq. (\ref{33}) are dominated by positive power function by analyzing these expressions and pictures of $Y(R)$, in which it can not find the form of inverted parabolic of $Y(R)$ out of the horizon . Thus the motion of  domain wall  can't realize the periodic oscillation outside of horizon. It shows that for the motion of domain wall, big bounce will not appear. On the other hand, this confirm that big bang theory is true again.
\section{the effect of dilaton field in the domain wall-bulk system}
Before this, it needs to introduce the  charged topological dilaton black hole solution gotten by Sheykhi \cite{Sheykhi:2007wg} for comparing the effect of dilaton field in different black hole model. The form of  metric is
\begin{equation}\label{bulkmetric}
ds^2=-U(r)dt^2+\frac{1}{U(r)}dr^2+r^2f^2(r)d\Omega_k^2,
\end{equation}
where $\Omega_k^2$ is the  line element of $n-2$ dimensional hypersurface and $k$ can take $0,1,-1$, respectively donated the horizon as flat, elliptic, hyperbolic.
\begin{eqnarray}
&&U(r)=-\frac{k(n-3)(a^{2}+1)^{2}b^{-2\gamma}r^{2\gamma}}{(a^{2}-1)(a^{2}+n-3)}-\frac{M}{r^{(n-2)(1-\gamma)-1}}\nonumber\\
&&~~~~~~+\frac{2Q^{2}(a^{2}+1)^{2}b^{-2(n-3)\gamma}}{(n-2)(a^{2}+n-3)}r^{2(n-3)(\gamma-1)}\nonumber\\
&&~~~~~~+\frac{2\Lambda(a^{2}+1)^{2}b^{2\gamma}}{(n-2)(a^{2}-n+1)}r^{2(1-\gamma)},\label{metricsolutionu}\\
&&\phi(r)=\frac{(n-2)a}{2(1+a^{2})}\ln(\frac{b}{r}),\label{potentialsolution}
\end{eqnarray}
where $b$ is a constant,  $\gamma=\frac{\alpha^{2}}{\alpha^{2}+1}$.
We  are going to use $R(r)$ to donate $rf(r)$ later
\begin{equation}
R(r)=rf(r)=b^{\frac{a^{2}}{(1+a^{2})}}r{}^{\frac{1}{(1+a^{2})}}.\label{rewritR}
\end{equation}

When $a=\gamma=0$, it can seen that this spacetime is an $n$ dimensional asymptotically AdS topological charged black hole. The metric have the form
\begin{eqnarray}
&&U(r)=k-\frac{M}{r^{n-3}}+\frac{2Q^{2}}{(n-2)(n-3)r^{2(n-3)}}\\\nonumber
&&~~~~~~~~~-\frac{2\Lambda}{(n-1)(n-2)}r^{2},\label{AdSmetricU}\\
&&R(r)=r,\label{AdSmetricR}
\end{eqnarray}
which is also studied by \cite{Brill:1997mf, Cai:1998vy}.

Here when $n=5$, we rewrite its  Friedmann equations
\begin{align}
&H^2=\frac{1}{36}\rho^2-\frac{R'^2}{R^2}U(r),\label{Hequation}\\
&\dot{\rho}+3H(P+\rho)=-\frac{4}{9}\rho\phi'^2H(\frac{R}{R'})^2\label{rhoequation}.
\end{align}
According to the Eqs. (\ref{potentialsolution}, \ref{rewritR}), the density is obtained as
\begin{equation}\label{density}
\rho=\rho_0+r^{-\frac{3(w+1)+a^2}{(1+a^2)}},
\end{equation}
where let $\rho_0=0$ for simplification. Then Eq. (\ref{Hequation}) can be simplified as
\begin{equation}\label{reHequation}
\frac{\dot{R}^2}{R^2}=\frac{1}{36}r^{\frac{2(3(w+1)+a^2)+2}{1+a^2}}
b^{\frac{2a^2}{1+a^2}}
-(\frac{1}{1+a^2})^2(\frac{r}{b})^{\frac{-2a^2}{1+a^2}}U(r).
\end{equation}
Combined with Eqs. (\ref{metricsolutionu}, \ref{rewritR},  \ref{reHequation}), $\gamma=\frac{\alpha^{2}}{\alpha^{2}+1}$ and a conversion $a=\sqrt{1-\beta}$, the dynamic equation can be
\begin{equation}\label{evolutionequation}
\begin{split}
\dot{R}^2&=Y(R)=\frac{1}{36}b^{\frac{-2}{\beta}
+2\beta}(b^{1-\beta})^{-\frac{2}{\beta}-\frac{6w}{\beta}}R^{-2-6w-2\beta}\\
&+\frac{b^{\frac{-4}{\beta}+\beta+\frac{2+\beta}{\beta}}
(b^{1-\beta})^{\frac{-1}{\beta}}m}{\beta^2}R^{-1-\beta}\\
&-\frac{2k}{2+\beta-\beta^2}-\frac{2b^{-4+\frac{2}{\beta}+2\beta}
(b^{1-\beta})^{\frac{-2}{\beta}}q^2}{3(\beta+\beta^2)}R^{-2-2\beta}\\
&\frac{2b^{2-\frac{4}{\beta}+2\beta}(b^{1-\beta})^{\frac{4}{\beta}}\Lambda}{3(-5+\beta)}
R^{4-2\beta}.
\end{split}
\end{equation}
Now let us make a simplified form, in which the monotonicity of Y(R) can be analyzed easily.
\begin{equation}\label{simplyYR}
\begin{split}
Y(R)&=R^{-2-6w-2\beta}-\frac{\Lambda}{-5-\beta}R^{4-2\beta}+R^{-1-\beta}
-R^{-2-\beta}\\
&-\frac{2k}{(2-\beta)(\beta+1)}.
\end{split}
\end{equation}

While $a=\gamma=0$, there is a evolution equation of the domain wall in  5 dimensional asymptotically AdS topological black hole
\begin{equation}\label{AdSevolutionequation}
\dot{R}^2=Y(R)=-k-\frac{Q^2}{3R^4}+\frac{1}{36}R^{-2(2+3w)}+MR^{-2}+R^2\frac{\Lambda}{6}.
\end{equation}
If $\Lambda<0$, it is very clear that it needs  $w<-1$. So that $Y(R)$ increases monotonously. If $\Lambda>0$, $Y(R)$ increases monotonously due to the term of $R^2$.

Now, comparing  the power of Eq. (\ref{AdSevolutionequation}) with the other expressions of $Y(R)$, it can be seen easily that all forms of power aimed  at the dilaton coupling constant is negative other than Eq. (\ref{33}). It can be inferred that the coupling between  dilaton field and Maxwell field  produce the effect of suppressive expansion in these spacetime. While Eq. (\ref{33}), due to the positive $\frac{4}{a^2}$ in the $C_1$, it shows that this coupling will stimulate expansion of domain wall. Thus the coupling between the dilaton field and Maxwell field is either suppressive or stimulative, for which, it depends on the different black hole model in the motion of domain wall.

Then if we choose the expansion stage as radiation stage $w=\frac{1}{3}$ or matter stage $w=0$, thus, we can pick the range of  the coupling strength among the dilaton field and Maxwell field. Because  astronomical data manifest that our universe is accelerating expansion \cite{Perlmutter:1998np}. Similarly, it needs to analyze the evolution of domain wall in the  charged topological dilaton black hole for comparing the coupling strength in the different black hole model. Thus  we do the following analysis.

Obviously, if $R>1$, $R^{-1-\beta}>R^{-2-2\beta}$. Thus the first three term of Eq.(\ref{simplyYR}) decides the monotonicity of $Y(R)$.

If $\beta>5$,  $4-2\beta<-1-\beta$. The last four terms will be dominated by $ R^{-1-\beta}$. $Y(R)$ will decrease monotonously. If$-2-6w-2\beta>0$, $w<-(1+\beta/3)$. $Y(R)$ will increase monotonously. If $-2-6w-2\beta < 0$, $w> -2$. $Y(R)$ will monotonously decrease.

If $1<\beta<5$, $4-2\beta>-1-\beta$.

(a) If $\Lambda>0$, the second term is big than 0. The last four terms are dominated by $-\frac{
\Lambda}{(-5+\beta)}R^{4-2\beta}$.

If $1<\beta<2$, $R^{4-2\beta}$ increases monotonously. If $-2-6w-2\beta>0$, $w<-1$. $Y(R)$ increases monotonously. If $-2-6w-2\beta<0$, $w>-\frac{2}{3}$. $Y(R)$ is dominated by the second term and it will increase monotonously.

If $2<\beta<5$, $R^{4-2\beta}$ decreases  monotonously. If  $-2-6w-2\beta<0$, $w>-1$. $Y(R)$ decreases monotonously. If $-2-6w-2\beta>0$, $w<-\frac{2}{3}$. $Y(R)$ increase monotonously.

(b)If $\Lambda<0$, the second term is litter than 0. If $Y(R)>0$, it needs $-2-6w-2\beta>4-2\beta$, namely, $w<-1$. If $w<-2$, $Y(R)$ increases monotonously; if $-\frac{1+\beta}{3}<w<-1$, $Y(R)$ will decreases monotonously.

Aimed at the trend that $Y(R)$ increases monotonously, there are three situation: case (I) $\beta>5, w<-\frac{1+\beta}{3}$. case(II) If $\Lambda>0$, $1<\beta<2, w<-1$ or $w>-\frac{2}{3}$; and  $2<\beta<5, w<-2$. case(III) $\Lambda<0$, $w<-2$.

Thus, When we choose the expansionary stage as radiation $w=\frac{1}{3}$ or matter $w=0$ stage, it just has one situation that  $\Lambda>0$, $1<\beta<2$, $w>-\frac{2}{3}$ for accelerating expansion. The range of dilaton coupling strength is $0<a<1$. However in the solution proposed by K. C. K. Chan et al , it can be seen that the value of  dilaton coupling constant $a$ can be chose as any domain from the Case (I) to Case (II) in the third   spacetime of the type II's solutions. For this, it infers that the coupling  strength between the dilaton field and Maxwell field  also depends on the selection of the metric. Moreover, For the rest of the metric solutions in this paper, it is obvious that they are not well. Because  when we chose the expansion stage as radiation $w=\frac{1}{3}$ or matter $w=0$ stage, the domain wall don't exist the accelerating expansion.
\section{Conclusion and Discussion}
In this paper,  the static bulk taken by K. C. K. Chan et al \cite{Chan:1995fr} are possessed of non-asymptotically flat and non-asymptotically (A)dS characteristics, under which it is expected to exhaust the motion of domain wall in the Einstein-Maxwell-dilaton system. By calculating, it shows that in different dilaton potentials domain wall always can cross horizon, for which  the  dilaton coupling constant $a$ and the ratio of pressure to density $w$ play the role of  controlling parameters. Aimed at expanding, it includes two situations. One is accelerating expansion and another is motion of constant velocity \cite{Mazharimousavi:2011km}. It can be seen from those pictures of $Y(R)$ that in fact, continuous expansion will be appeared very early and sharply. Unless we change the others parameters i.e.,$\Lambda, \gamma, M, Q$, it will produce the result of FIG. (\ref{11picture}).  Beyond that, for the expansion we can simply analyze the stability of  domain wall by calculating $Y''(R)$, for which it learns from analyzing the stability of wormhole \cite{Eiroa:2005pc}-\cite{Setare:2014eba}. Obvious, when domain wall exist matter,  $Y''(R)>0$ can be found from these pictures and calculation of $Y(R)$. In this way, these walls are simply deemed to be not stability   in all the situations of expansion. Further more, our analysis reveals that  the big bang theory applies to the domain wall world scenario, while big bounce will not appear in our paper.

In addition to this, it can be found that  the coupling between  dilaton field and Maxwell field  is either suppressive or stimulative for the expansion of domain wall, by comparing the power of  dynamic equation. It shows that the effect of coupling depends on the model. When we choose the expansion stage as radiation $w=\frac{1}{3}$ or matter $w=0$ stage, the resultant coupling strength   between  dilaton field and Maxwell field also depends on the black hole models. At the same time, we find some spacetime is not well according to the fact, our universe is  accelerating expansion.

For extensional work, Firstly  we are interested in the perturbation of domain wall in this system. We are desired that if can the coupling strength be confirmed by analyzing the stability of domain wall. Secondly it is desired that if this expansion can be considered as inflation? In case, ending inflation can be come true by altering some parameters. Thirdly, we can study the hybrid inflation. And for it, Higgs field can be inflaton  and  dilaton field is used to end inflation. Finally in our next paper we plan to discuss the big bounce or big bang on the domain wall in other model when charged black hole bulk action adds dilaton field \cite{Medved:2002mi}.
\section*{Acknowledgements}
The work is supported by National Natural Science Foundation of China (No.11875081 and No.10875009).

\end{document}